\documentclass[final,1p,times]{elsarticle}

\usepackage{amssymb}
\usepackage{amsmath}
\usepackage{graphicx}
\usepackage{xcolor}
\usepackage{bm}
\usepackage{algorithm}
\usepackage{algpseudocode}
\usepackage{soul}
\usepackage{url}
\usepackage{natbib}
\journal{Computer Physics Communications}

\begin{document}

\begin{frontmatter}

\title{Accelerating multigrid with streaming chiral SVD for Wilson fermions in lattice QCD}

\author{Travis Whyte} 

\affiliation{organization={J\"ulich Supercomputing Center},
            city={J\"ulich 52428},
            country={Germany}}

\author{Andreas Stathopoulos} 

\affiliation{organization={Department of Computer Science, William \& Mary},
            city={Williamsburg},
	    state={Virginia},
            country={USA}}

\author{Eloy Romero} 

\affiliation{organization={Thomas Jefferson National Accelerator Facility},
            city={Newport News},
            state={Virginia},
            country={USA}}

\begin{abstract}
A modification to the setup algorithm for the multigrid preconditioner of Wilson fermions in lattice QCD is presented. A larger basis of test vectors than that used in conventional multigrid is calculated by the smoother and truncated by singular value decomposition on the chiral components of the test vectors. The truncated basis is used to form the prolongation and restriction matrices of the multigrid hierarchy. This modification of the setup method is demonstrated to increase the convergence of linear solvers on an anisotropic lattice with $m_{\pi} \approx 239$ MeV from the Hadron Spectrum Collaboration and an isotropic lattice with $m_{\pi} \approx 220$ MeV from the MILC Collaboration. The lattice volume dependence of the method is also examined. Increasing the number of test vectors improves speedup up to a point, but storing these vectors becomes impossible in limited memory resources such as GPUs. To address storage cost, we implement a \emph{streaming} singular value decomposition of the basis of test vectors on the chiral components and demonstrate a decrease in the number of fine level iterations by a factor of 1.7 for $m_q \approx m_{crit}$.
\end{abstract}


\end{frontmatter}

\section{Introduction}
\label{intro}

Quantum Chromodynamics (QCD) is the theory of the strong interaction, which governs the dynamics of quarks and gluons. A non-perturbative approach to QCD is \textit{lattice} QCD, wherein the equations governing the strong interaction are discretized to a hypercubic spacetime lattice where they can be solved numerically. One of the most computationally expensive parts of any lattice QCD calculation is the generation of \textit{propagators}, which are created by solving systems of linear equations 
\begin{equation}
\label{eq::linear}
	\bm{D}x = b
\end{equation}
where $\bm{D} \in \mathbb{C}^{n \times n}$ is the lattice Dirac operator, a large sparse matrix arising from the discretization of the quark and gluon fields. This matrix has a dependence on the underlying background gauge field, quark mass and volume of the spacetime lattice. As the mass of the light quarks approaches their physical value, the eigenspectrum of the lattice Dirac operator approaches the origin of the complex plane. In this regime, iterative solvers experience \textit{critical slowing-down}, wherein the number of iterations required for the system of linear equations to reach convergence begins to drastically increase as the lowest eigenvalues of the matrix approach zero. Critical slowing-down is also experienced as the volume of the lattice grows larger, which results in increased density of the matrix spectrum near zero. This increased density near the origin of the complex plane further increases the difficulty of solving the system of linear equations, and preconditioning methods of the iterative solvers are required. 

Many preconditioning methods have been developed and used within lattice QCD to combat the increasing computational cost of generating propagators. Such methods include even-odd preconditioning, ILU, additive and multiplicative Schwartz, deflation, polynomial preconditioning and most recently, multigrid; see a good introduction to these techniques at \cite{saad2003iterative}. In the first multigrid attempts for lattice QCD, a renormalization group approach was utilized to construct the multigrid hierarchy (see a review of early work at \cite{KALKREUTER199557}). While the multigrid preconditioner created in this approach exhibited speedup compared to conventional methods, critical slowing down was not fully eliminated on the finest grid. It was not until the advent of adaptive multigrid \cite{Brannick2008} that a successful multigrid preconditioner for Wilson fermions was created. In adaptive multigrid, the components of the error that the smoother cannot sufficiently reduce, i.e. the low frequency components, are exposed by using the smoother to create a set of algebraically smooth near null or test vectors. These test vectors are rich in components in the direction of the low-lying eigenspace and are used to create a coarse grid matrix through a projection of the original fine grid matrix. This process can be performed recursively to create matrices associated with coarser grids. 

The number of test vectors used to create the coarse grid matrices must be chosen judiciously. A large number of test vectors results in coarse grid matrices that are efficient at reducing the error associated with the low lying eigenspace but are expensive to apply. This is because the degrees of freedom at each grid point of the coarser grid, and by extension the number of non-zeros of the matrix, scale quadratically with the number of test vectors. Consequently, there is an overall increase in the time to solution due to the cost of the stencil application and the time spent communicating the enlarged halos \cite{clark2016acceleratinglatticeqcdmultigrid}. 
Conversely, having too few test vectors leads to small coarse grid matrices that are inexpensive to apply. However, the resulting preconditioner may not adequately reduce the error associated with the low-lying eigenspace, which results in slower convergence. We are motivated in looking for an improvement in the multigrid setup process that can achieve the convergence properties of a multigrid preconditioner created with a large number of test vectors while maintaining the cost of one constructed with a smaller number of test vectors.

Since Ref \cite{Brannick2008}, multigrid preconditioners for other fermion discretizations have been created, eg. \cite{Brower2018,Brower2020}. However, little work has been done to improve the multigrid setup process for Wilson fermions beyond its initial implementation. Most studies have focused on specific parts of the setup process, such as smoothers \cite{brannick2016multigrid,frommer2014adaptive}, reducing the condition number of the coarse grid matrix \cite{ESPINOZAVALVERDE2023108869}, or replacing the multigrid preconditioner with a deep neural network \cite{lehner2023}. In this article, a modification of the setup method for the multigrid preconditioner of Wilson type fermions is presented, which improves the basis of test vectors used to create the coarse grid matrices through a singular value decomposition and subsequent truncation of the basis. In Section \ref{sec:background}, the discretization for Wilson fermions and their multigrid setup method are reviewed. In Section \ref{sec:rankk}, we introduce the singular value decomposition setup method, the modifications for Wilson type fermions and the \emph{streaming} method used to address the storage cost of the enlarged basis of test vectors. Section \ref{sec:exp} demonstrates the applicability of the method on two separate gauge configurations from the MILC Collaboration and the Hadron Spectrum Collaboration and examines the volume dependence, as well as the performance of the streaming method. In Section \ref{sec:summ}, we summarize our results and provide future avenues of research.

\section{Background}
\label{sec:background}

\subsection{Multigrid for Wilson Type Fermions}
\label{subsec:mg_wilson}

Classical multigrid fails due to the near random nature of the $SU(3)$ gauge links. 
Only with the advent of adaptive multigrid was it possible to obtain a preconditioner for the Wilson-Dirac operator that significantly mitigated critical slowing down of the solver, that is, the number of iterations was increasing more slowly as the operator mass approached the critical mass.
Under adaptive multigrid, the smoother is tested on the homogeneous equation $\bm{D}\psi_i \approx 0$ to create a set of $k$ near null or test vectors, $\bm{\Psi} = [\psi_1,\dots,\psi_k]$, with random initial guesses, $\phi_i$. That is equivalent to do
\begin{equation}
	 \psi_i=\phi_i+\psi_i' \quad \text{where\ } \bm{D}\psi_i' \approx -\bm{D}\phi_i,\qquad \text{for\ } i = 1, \dots, k.
\end{equation}
This procedure generates $\psi_i$ rich in components of the error that the smoother is not able to easily resolve, i.e. the low frequency components of the error. The test vectors generated display the phenomenon known as \textit{local coherence}, wherein the smooth test vectors approximate the low lying eigenspace of $\bm{D}$ on a given domain \cite{Luscher2007,Luscher2007a}. Thus, in the smooth aggregation approach, these smooth test vectors are then blocked according to a domain decomposition of the underlying grid, which means that each test vector is split into multiple vectors, each having support only on one of the subdomains. These vectors then become the columns of the \textit{prolongator} matrix, $\bm{P}$. The \textit{restriction} matrix, $\bm{R}$, is typically taken to be $\bm{R} = \bm{P}^{\dag}$. However, the success of the multigrid preconditioner for the Wilson-Dirac operator relies on ``chiral splitting'', wherein the smooth test vectors are further split into chiral components by applying the projectors $1 \pm \gamma_5$ which split the spin components of $\bm{\Psi}$ such that
\begin{align}
	\nonumber \bm{\Psi}^{(\alpha = 0)} & = [\psi_1^{(\beta = 1,2)},\dots,\psi_k^{(\beta = 1,2)}] \\
	\bm{\Psi}^{(\alpha = 1)} & = [\psi_1^{(\beta = 3,4)},\dots,\psi_k^{(\beta = 3,4)}],
	\label{eq::chiral_split}
\end{align}
where $\alpha$ labels the chiral index and $\beta$ labels the spin index of the test vectors. Color indices have been suppressed for brevity. The blocked and chirally-split test vectors are then orthonormalized on each domain of the spacetime lattice $\Lambda_j \in \Lambda = \{\Lambda_1,\dots,\Lambda_d\}$ and the prolongator is formed as
    \begin{equation}
	    P_{\alpha ij}(x) = \begin{cases} 
      \psi^{(\alpha)}_i(x) & \mathrm{if}~x\in \Lambda_j \\
      0 & \mathrm{otherwise}
   \end{cases},
    \label{eq:prolong}
    \end{equation}
where $x$ denotes the lattice coordinate. This results in a block diagonal matrix of dimension $n \times 2kd$, where $d = |\Lambda|$. The use of the projectors $1 \pm \gamma_5$ enforces the symmetry $\gamma_5 \bm{P} = \bm{P} \sigma_3$, where $\sigma_3$ is the third Pauli spin matrix, and ensures that if a right singular vector is in the range of $\bm{P}$, a left singular vector is in the range of $\bm{R}$. The coarse grid matrix is then formed via
\begin{equation}
	\bm{D}_{\ell+1} = \bm{RD}_{\ell}\bm{P}.
\end{equation}
This process is repeated recursively beginning with the coarse grid operator $D_{\ell+1}$ to form successively coarser grids. The projector $(1\pm \gamma_5^{\ell})$ is used to perform the chiral splitting, where $\gamma_5^{\ell} = \sigma_3$ for $\ell > 0$. The multigrid setup process is summarized in Algorithm \ref{alg::mg}. Typical applications of multigrid in lattice QCD utilize the multigrid hierarchy in a $K$-cycle \cite{Notay2008} as a preconditioner for a flexible outer solver such as FGMRES or GCR.
\begin{algorithm}
	\caption{The multigrid setup algorithm for Wilson fermions.}
	\label{alg::mg}
	\textbf{Input}: $\bm{D}_{\ell}$, $\Lambda$, $d$, $k$, $\ell_{max}$\\
	\textbf{Output}: $\bm{D}_{\ell+1}$, $\bm{R}$, $\bm{P}$
        \begin{algorithmic}[1]
	\For{$\ell = 0,\dots,\ell_{max}-1$}
		\For{$i = 1,\dots,k$} \\
			\hspace*{\algorithmicindent}\hspace*{\algorithmicindent}Smooth on $\bm{D}_{\ell}\psi_i \approx 0$ with random initial guess
		\EndFor \\
		\hspace*{\algorithmicindent}Apply the projectors $(1\pm\gamma_5^{\ell})$ to $\bm{\Psi} = [\psi_1,\dots,\psi_k]$
		\For{$\alpha = 0,1$} 
			\For{$j = 1,\dots,d$} \\
				\hspace*{\algorithmicindent}\hspace*{\algorithmicindent}\hspace*{\algorithmicindent}Orthonormalize $\bm{\Psi}^{(\alpha)}(\Lambda_j)$
			\EndFor
		\EndFor \\
		\hspace*{\algorithmicindent}Form $\bm{P}$ as in Equation (\ref{eq:prolong}); take $\bm{R} = \bm{P}^{\dag}$\\
		\hspace*{\algorithmicindent}Form $\bm{D}_{\ell+1} = \bm{R}\bm{D}_{\ell}\bm{P}$
	\EndFor
	\end{algorithmic}
\end{algorithm}

\section{Chiral SVD of the Multigrid Basis}
\label{sec:rankk}
As mentioned earlier, a larger number of test vectors captures better the near null space providing a more effective preconditioner in terms of multigrid iterations. However, the size of the coarse grid matrix ($2kd$) also increases and thus it is not beneficial to increase $k$ beyond a certain value\footnote{a value of $k$ around 24 is often used for the first level, and between 32 and 96 are often used for the subsequent levels.}. This tradeoff in smooth aggregation behavior is also common in other fields, and various approaches have been proposed to improve the quality of the test vector basis \cite{dambra2019,Chow2006,doi:10.1137/090752973}. 

One such approach finds $m>k$ smooth test vectors and truncates them to their best rank-$k$ approximation using the singular value decomposition. Using the rank-$k$ approximation as the multigrid basis of smooth test vectors was given in \cite{Chow2006} and was shown to increase the performance of the multigrid preconditioner for two-dimensional anisotropic heat diffusion equations. It was also used in Ref. \cite{dambra2019} to form composite aggregate preconditioners in bootstrap AMG. We review the rank-$k$ approximation method given in \cite{Chow2006} and then specify the modifications needed for Wilson fermions in lattice QCD. 

Without loss of generality, we briefly assume only one degree of freedom per site (that is, every lattice site is then described by a single number) on a lattice of volume $V$ and a uniform domain decomposition of the lattice where each domain $\Lambda_j$ contains $n_d$ lattice sites. Given a set of $m > k$ smooth test vectors at level $\ell$, $\bm{\Psi}^{(\ell)} = [\psi_1,...,\psi_m]$, 
consider its singular value decomposition restricted to a particular domain $\Lambda_j$,  $ \bm{\Psi}^{(\ell)}(\Lambda_j) = \bm{U}\bm{\Sigma}\bm{V}^{\dag}$. Let $\bm{U}_k \in \mathbb{C}^{n_d \times k}$ be the first $k$ left singular vectors, $\bm{\Sigma}_k \in \mathbb{R}^{k \times k}$ the first $k$ singular values and $\bm{V}_k \in \mathbb{C}^{m \times k}$ the first $k$ right singular vectors. Then, the best rank-$k$ approximation to $\bm{\Psi}^{(\ell)}(\Lambda_j)$ in terms of the 2-norm error is $\bm{U}_k\bm{\Sigma}_k\bm{V}_k^{\dag} \approx \bm{\Psi}^{(\ell)}(\Lambda_j) $. Then we can use the rank-$k$ left singular vector basis for this approximation to obtain the prolongator for the next level, i.e., setting $\bm{P}_j = \bm{U}_k$ for each domain.  

The extension of this method to Wilson fermions in lattice QCD is straightforward. The lattice sites are described by twelve degrees of freedom: 4 from the spin components and 3 from the color components. Guided by the required chiral splitting in the conventional set up method, the singular value decomposition is performed separately on the chiral components of the initial set of $m$ smooth test vectors given by (\ref{eq::chiral_split}), resulting in six degrees of freedom per lattice site for each chiral index. Let $\bm{U}^{(\alpha)}\bm{\Sigma}^{(\alpha)}\bm{V}^{^{(\alpha)}\dag} = \bm{\Psi}^{(\alpha,\ell)}(\Lambda_j)$ be the singular value decomposition of $\bm{\Psi}^{(\alpha,\ell)}(\Lambda_j)$. 
Then, for $\alpha = 0,1$, the prolongator is formed as
    \begin{equation}
            P_{\alpha ij}(x) = \begin{cases}
		    u^{(\alpha)}_i(x) & \mathrm{if}~x\in \Lambda_j \\
      0 & \mathrm{otherwise}
   \end{cases}.
    \label{eq::chiral_prolong}
    \end{equation}
As the singular value decomposition creates a basis of $m$ orthonormal left singular vectors for each chiral index, the orthonormalization of step $8$ in Algorithm \ref{alg::mg} is not needed.
We summarize the full multigrid set up algorithm utilizing the chiral SVD (CSVD) in Algorithm \ref{alg::cmg}.

It is apparent that the cost of creating the multigrid preconditioner using Algorithm \ref{alg::cmg} scales as $\mathcal{O}(m)$ in the number of linear systems to be smoothed and incurs also a much smaller $O(nm^2)$ SVD cost. Thus, it is more expensive and requires more storage than Algorithm \ref{alg::mg}. However, most applications in lattice QCD require solving Equation (\ref{eq::linear}) many times. Spectroscopy, which uses the distillation method \cite{peardon2009}, requires Equation (\ref{eq::linear}) be solved with $\mathcal{O}(10^4)$ right hand sides. In such a situation, the increased cost of generating more vectors would be heavily amortized. In the following section, we address the cost of storage through a streaming SVD, or incremental SVD (iSVD) method. 
\begin{algorithm}
        \caption{The multigrid setup algorithm for Wilson fermions with CSVD.}
        \label{alg::cmg}
        \textbf{Input}: $\bm{D}_{\ell}$, $\Lambda$, $d$, $m$, $k$, $\ell_{max}$\\
        \textbf{Output}: $\bm{D}_{\ell+1}$, $\bm{R}$, $\bm{P}$
        \begin{algorithmic}[1]
	\Require $m \geq k$
	\For{$\ell = 0,\dots,\ell_{max}-1$}
                \For{$i = 1,\dots,m$} \\
			\hspace*{\algorithmicindent}\hspace*{\algorithmicindent}Smooth on $\bm{D}_{\ell}\psi_i \approx 0$ with random initial guess
                \EndFor \\
		\hspace*{\algorithmicindent}Apply the projectors $(1\pm\gamma_5^{\ell})$ to $\bm{\Psi} = [\psi_1,\dots,\psi_m]$
                \For{$\alpha = 0,1$}
                        \For{$j = 1,\dots,d$} \\
				\hspace*{\algorithmicindent}\hspace*{\algorithmicindent}\hspace*{\algorithmicindent}$[\bm{U}_k^{(\alpha)},\bm{\Sigma}_k^{(\alpha)},\bm{V}_k^{(\alpha)}] = \text{rank-}k\ \mathrm{SVD}(\bm{\Psi}^{\alpha}(\Lambda_j))$
                        \EndFor
                \EndFor \\
		\hspace*{\algorithmicindent}Form $\bm{P}$ as in Equation (\ref{eq::chiral_prolong}); take $\bm{R} = \bm{P}^{\dag}$\\
		\hspace*{\algorithmicindent}Form $\bm{D}_{\ell+1} = \bm{R}\bm{D}_{\ell}\bm{P}$
	\EndFor
        \end{algorithmic}
\end{algorithm}

\subsection{Streaming CSVD of the Basis}
The storage cost of the CSVD method in the proceeding section scales linearly with $m$ for a fixed domain size. A problem arises when it is necessary to choose a large value of $m$ to obtain optimal multigrid performance (as shown in Section \ref{sec:exp}), but the computational platform has memory-constrained resources, such as GPUs, and therefore not all test vectors can be stored to compute the SVD. To address this, we implement an iSVD \cite{brand2002incremental} method, which incrementally calculates approximate singular triplets of a matrix that is not available in its entirety but is streamed in windows of columns.

To facilitate the discussion of iSVD, we define the matrix
\begin{equation}
    \bm{\Phi}_k^{(\alpha)}(\Lambda_j) = \bm{U}_k^{(\alpha)}\bm{\Sigma}_k^{(\alpha)},
    \label{eq::phi}
\end{equation}
where $\bm{U}_k^{(\alpha)}$ and $\bm{\Sigma}_k^{(\alpha)}$ are the first $k$ left singular vectors and singular values of $\bm{\Psi}^{\alpha}(\Lambda_j)$. Let each stream be labeled by subscript $s$, the number of streams be represented by $n_s$, and the size of the initial basis $m = k + r$, where $r$ is the number of test vectors generated at each stream. At the first stream, iSVD does not have any prior rank-$k$ approximation, so
$\bm{\Phi}_k^{(\alpha)}(\Lambda_j)_0 = \emptyset$. For each stream then, iSVD updates the approximation as
\begin{equation}
    \bm{\Phi}_k^{(\alpha)}(\Lambda_j)_{s}= \text{rank-}k~\text{SVD}([\bm{\Phi}_k^{(\alpha)}(\Lambda_j)_{s-1},~~~\bm{\Psi}_r^{(\alpha)}(\Lambda_j)_s])
    \label{eq::phi_update}
\end{equation}
where $\bm{\Psi}^{(\alpha)}(\Lambda_j)_s$ are the $r$ newly generated test vectors during the stream $s$, which are appended to $\bm{\Phi}_k^{(\alpha)}(\Lambda_j)_{s-1}$. During each stream, the SVD is performed on Equation (\ref{eq::phi_update}) and the first $k$ left singular vectors and singular values are kept as in Equation (\ref{eq::phi}). 

\begin{algorithm}
        \caption{The iCSVD algorithm.}
        \label{alg::stream_svd}
        \textbf{Input}: $\bm{D}_{\ell}$, $\Lambda$, $d$, $m$, $k$, $r$, $n_s$
        \textbf{Output}: $U_k$
        \begin{algorithmic}[1]
	\Require $m = k + r$. Initially $m = r$ and $\bm{\Phi}_m^{(\alpha)}(\Lambda_j)_{-1}=\emptyset$.
	\For{$s = 0,\dots,n_s-1$}
                \For{$i = 1,\dots,r$} \\
\hspace*{\algorithmicindent}\hspace*{\algorithmicindent}Smooth on $\bm{D}_{\ell}\psi_i \approx 0$ with random initial guess
                \EndFor \\
		\hspace*{\algorithmicindent}Apply the projectors $(1\pm\gamma_5^{\ell})$ to $\bm{\Psi}_s = [\psi_1,\dots,\psi_r]$
                \For{$\alpha = 0,1$} \\
                        \hspace*{\algorithmicindent}\hspace*{\algorithmicindent}$\bm{\Phi}_m^{(\alpha)}(\Lambda_j)_s = [\bm{\Phi}_k^{(\alpha)}(\Lambda_j)_{s-1},~~\bm{\Psi}^{(\alpha)}(\Lambda_j)_s]$
                        \For{$j = 1,\dots,d$} \\
				\hspace*{\algorithmicindent}\hspace*{\algorithmicindent}\hspace*{\algorithmicindent}$[\bm{U}_k^{(\alpha)},\bm{\Sigma}_k^{(\alpha)},\bm{V}_k^{(\alpha)}] = \text{rank-}k\ \mathrm{SVD}(\bm{\Phi}_m^{(\alpha)}(\Lambda_j)_s)$ 

                \If{$s == n_s - 1$} \\
                \hspace*{\algorithmicindent}\hspace*{\algorithmicindent}\hspace*{\algorithmicindent}\hspace*{\algorithmicindent}return $\bm{U}_k$
                \Else \\
                \hspace*{\algorithmicindent}\hspace*{\algorithmicindent}\hspace*{\algorithmicindent}\hspace*{\algorithmicindent}$\bm{\Phi}_k^{(\alpha)}(\Lambda_j)_s = \bm{U}_k^{(\alpha)}\bm{\Sigma}_k^{(\alpha)}$ 
                \EndIf
                        \EndFor
                \EndFor
    \EndFor
\end{algorithmic}
\end{algorithm}

Algorithm \ref{alg::stream_svd} gives our implementation of the iSVD algorithm, which accounts for the chirality of the test vectors. We therefore refer to it as incremental chiral SVD (iCSVD). 
Streaming SVD does not compute the singular vectors as accurately as SVD on the full set of vectors. However, its accuracy is typically higher than what is needed to build multigrid prolongators. Moreover, iCSVD allows us to sample a larger number of test vectors incrementally, keeping only the first $k$ left singular vectors and singular values for each stream, resulting in a total space complexity of $\mathcal{O}(nm)$. In contrast, the non-streaming version of this method would need to store $n_sm$ test vectors for a comparable sample, resulting in a total space complexity of $\mathcal{O}(nn_sm)$. The iCSVD then allows us to sample more test vectors without the increase in storage cost and should result in a more effective preconditioner. Additionally, the use of iCSVD slightly reduces the total time complexity, which scales as $\mathcal{O}(nn_sm^2)$, compared to a single SVD on all vectors. For example, if the number of test vectors used for the streaming version is $m$, and the number of test vectors for the non-streaming version is $n_sm$, then an equal number of test vectors have been sampled between the two methods. However, the time complexities between the two methods will differ by a factor of $n_s$, leading to a speedup in the total time spent performing the SVD by a factor of $\mathcal{O}(n_s)$. Algorithm \ref{alg::csmg} summarizes the the full multigrid setup algorithm with the iCSVD method. 

\begin{algorithm}
        \caption{The multigrid setup algorithm for Wilson fermions with iCSVD.}
        \label{alg::csmg}
        \textbf{Input}: $\bm{D}_{\ell}$, $\Lambda$, $d$, $m$, $k$, $r$, $n_s$, $\ell_{max}$\\
        \textbf{Output}: $\bm{D}_{\ell+1}$, $\bm{R}$, $\bm{P}$
        \begin{algorithmic}[1]
	\Require $m = k + r$
	\For{$\ell = 0,\dots,\ell_{max}-1$} \\
         \hspace*{\algorithmicindent}$\bm{U}_k =$ iCSVD($\bm{D}_{\ell}$, $\Lambda$, $d$, $m$, $k$, $r$, $n_s$); \\     
		\hspace*{\algorithmicindent}Form $\bm{P}$ as in Equation (\ref{eq::chiral_prolong}); take $\bm{R} = \bm{P}^{\dag}$\\
		\hspace*{\algorithmicindent}Form $\bm{D}_{\ell+1} = \bm{R}\bm{D}_{\ell}\bm{P}$
	\EndFor
        \end{algorithmic}
\end{algorithm}

\section{Numerical Experiments}
\label{sec::exp}
The setup method with CSVD is tested against conventional multigrid for two different types of lattice configurations. The first, which we refer to as Ensemble A, corresponds to an anisotropic lattice of dimension $32^3 \times 256$ with a pion of mass $m_{\pi} \approx 239$ MeV from the Hadron Spectrum Collaboration \cite{Lin2009,edwards2008tuning}. The gauge links arising from these configurations are stout smeared \cite{morningstar2004analytic}. See \cite{wilson2019} for more information relating to these lattices, including scale and mass determination. The second, which we refer to as Configuration B, corresponds to an isotropic lattice of dimension $32^3 \times 64$ with a pion of mass $m_\pi \approx 220$ MeV from the MILC Collaboration \cite{milc0,milc1}. In this Clover on HISQ action, the gauge links are smeared with one iteration of HYP smearing \cite{hyp}. Ref. \cite{Gupta2018} contains more information on the determination of the relevant physical properties of this ensemble. All experiments were performed on eight nodes with Intel Xeon Gold 6130 processors having $2 \times 16$ cores. In all experiments, the results are averaged over ten right hand sides and FGMRES was used as both a smoother and solver for each level. Unless explicitly stated otherwise, 8 iterations of a post-smoother were utilized on levels $\ell = 0,1$ and the linear systems on levels $\ell = 1,2$ were solved to a residual tolerance of $||r_\ell||_2 = 0.03||b_\ell||_2$. 
\label{sec:exp}

\subsection{Size of the Multigrid Basis}
\label{subsec:size}
As the size of the initial multigrid basis $m$ is increased, it is natural to expect that more information about the near null space is captured by the preconditioner, leading to an increase in convergence of the linear equations. Typical applications of multigrid in lattice QCD use between 24 and 32 test vectors. However, the CSVD allows us to calculate an initial basis size of $m \geq k$ and incorporate information from the test vectors that the conventional setup method ignores. It is thus important to quantify the performance increase offered by increasing $m$.
\begin{figure}[!h]
	\begin{center}
	\includegraphics[scale=0.525]{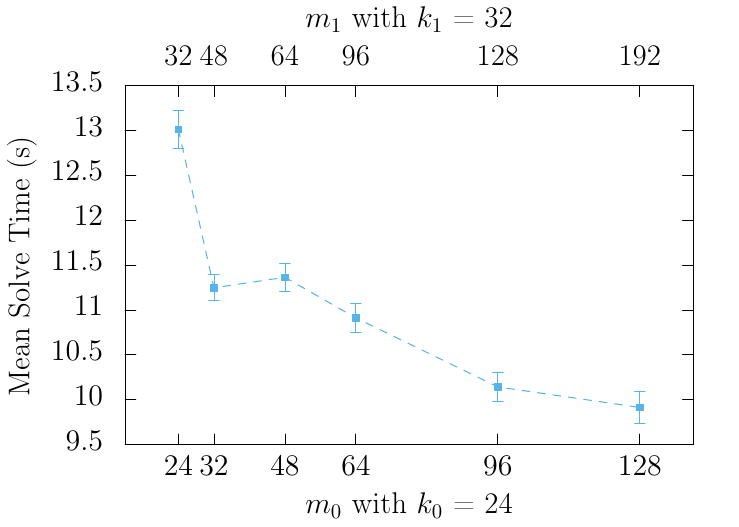}
	\includegraphics[scale=0.525]{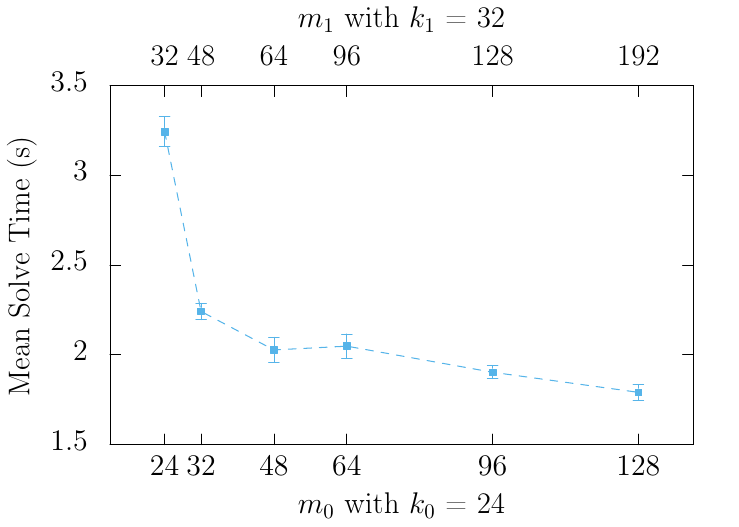}
		\caption{\label{fig::basissize}The mean solve time as a function of the initial basis size for Configuration A (left) and Configuration B (right). The bottom and top $y$-axis displays $m$ for levels $\ell = 0,1$, respectively.}
	\end{center}
\end{figure}
Figure \ref{fig::basissize} shows the mean execution time for solving the systems of linear equations for both Configuration A and Configuration B. The basis is truncated at $k = 24,32$ for levels $\ell = 0,1$, respectively. A general trend is observed for both configurations: as $m$ increases, the mean execution time decreases. This is in line with expectations that additional information from the exposed smooth modes is made available to the preconditioner. It is also observed that a saturation in the speedup begins between $m = 96$ and $m = 128$, indicating that beyond $m = 96$ the rate of new near-null information in test vectors reduces and the quality of the prolongators stabilizes. 

\subsection{Optimal Truncation}
\label{subsec:opt_trunc}
The number of test vectors used to create the prolongation and restriction matrices is an important determination. Too few vectors results in a poor preconditioner and too many test vectors results in large and dense coarse grid matrices that are expensive to apply. It is thus important to examine the rank of the truncation utilized.  
\begin{figure}[!h]
	\begin{center}
	\includegraphics[scale=0.525]{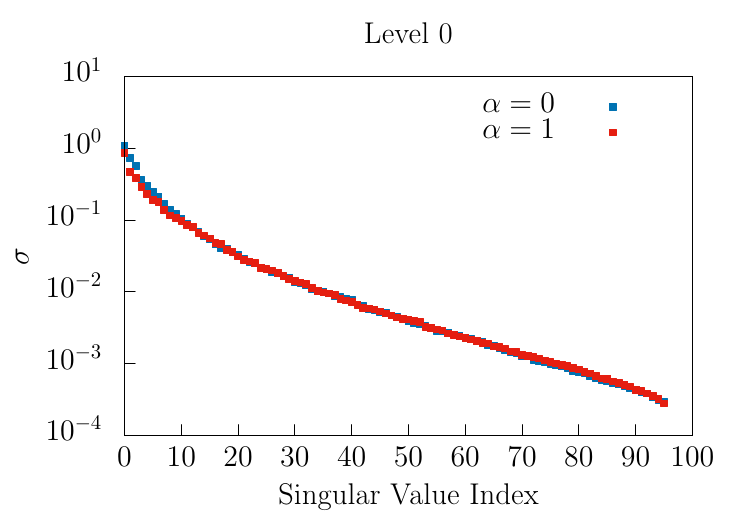}
	\includegraphics[scale=0.525]{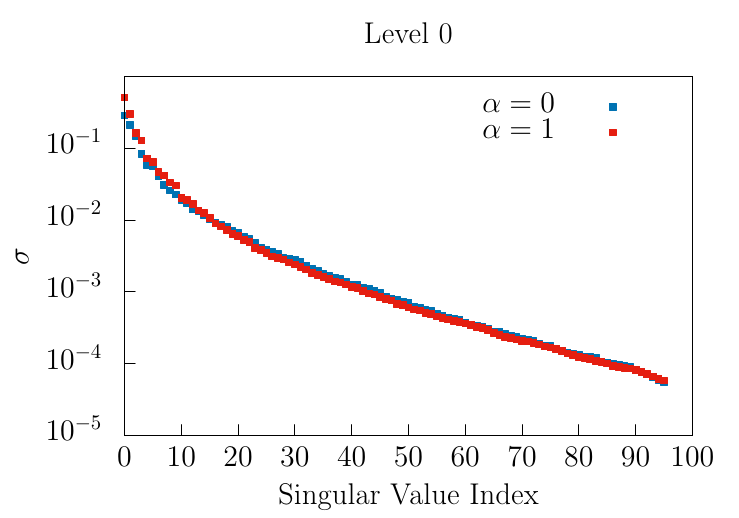}
	\includegraphics[scale=0.525]{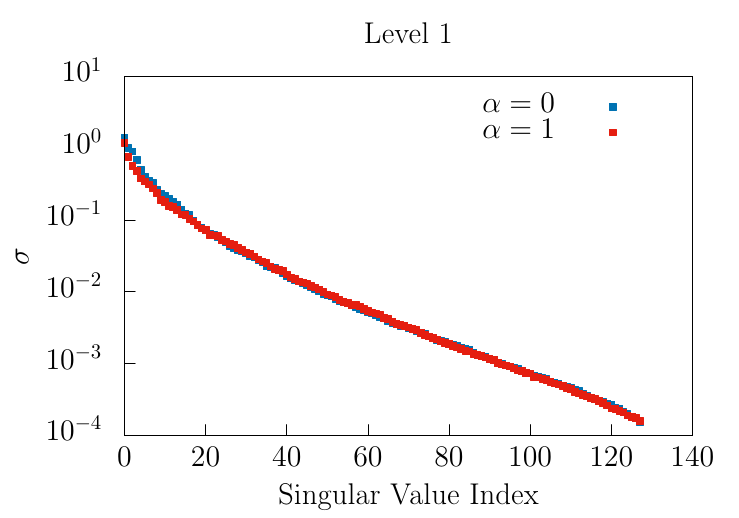}
	\includegraphics[scale=0.525]{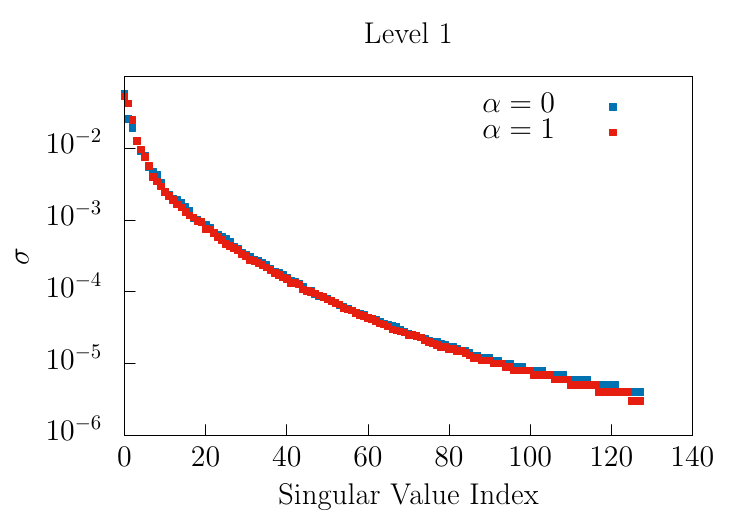}
		\caption{\label{fig::svs}The singular spectrum for the chirally split test vectors on the first domain for level $\ell = 0$ (top) and $\ell = 1$ (bottom) of Configuration A (left) and Configuration B (right).}
	\end{center}
\end{figure}
The singular spectra of the test vectors provides some information for this determination. Figure \ref{fig::svs} shows the singular spectrum of the chirally split test vectors for the first domain, which is representative of other domains. It is observed that the spectrum for Configuration A displays superlinear decay for approximately the first 16 singular values, and the spectrum for Configuration B displays superlinear decay for approximately the first 24 singular values for $\ell = 0$ for both chiral components. After these respective values, the spectrum displays linear decay. A similar observation is made for the level $\ell = 1$, with linear decay of the spectrum arising approximately after the first 30 singular values. The transition from superlinear decay to linear decay indicates that including left singular vectors into the basis beyond this approximate transition point will have a smaller overall contribution to the low rank approximation of the multigrid basis. 

\begin{figure}[!h]
	\begin{center}
	\includegraphics[scale=0.525]{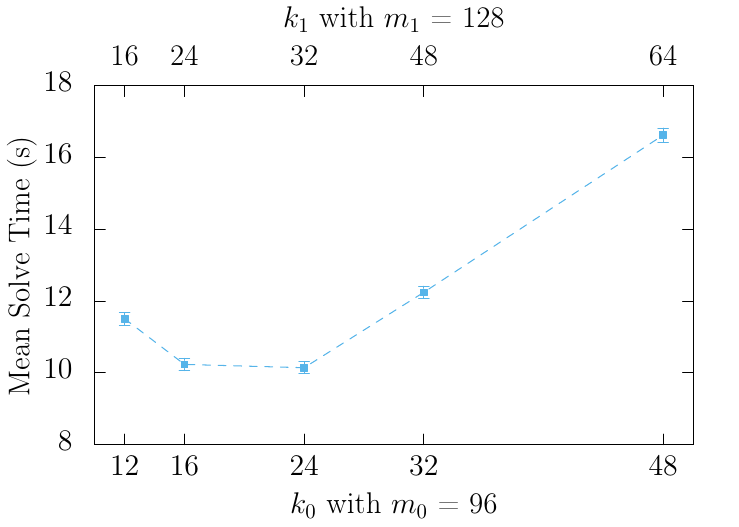}
	\includegraphics[scale=0.525]{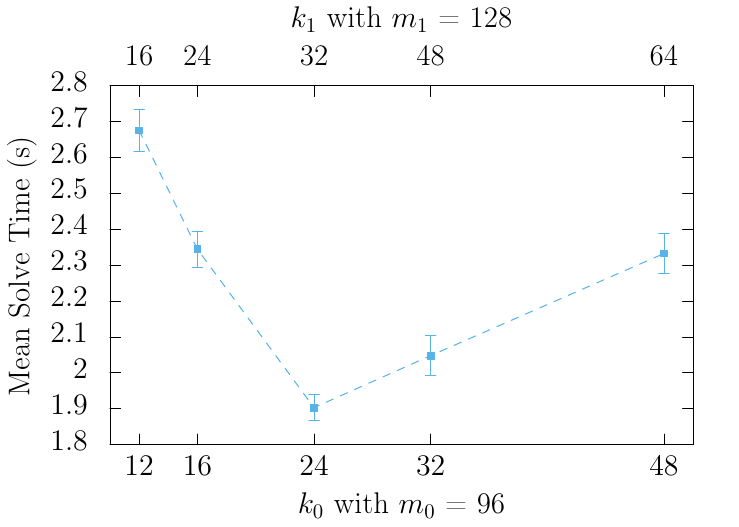}
		\caption{\label{fig::rank}The mean execution time for the system of linear equations when the degree of the truncation is varied using an initial basis size of $m = 96$ for Configuration A (left) and Configuration B (right). The bottom and top $y$-axis displays $k$ for levels $\ell = 0,1$, respectively.}
	\end{center}
\end{figure}

Figure \ref{fig::rank} displays the mean execution time of the system of linear equations when the truncation size, $k$, is varied for a constant value of $m = 96$ (as this $m$ achieves most of the reduction in execution time). For both configurations, the optimal truncation is observed at $k = 24$. For Configuration A, this is comparable to $k = 16$, while Configuration B exhibits a distinct minimum at $k = 24$. 

\subsection{Optimal Number of Setup Iterations}
\label{subsec:setup}
The number of set up iterations of the smoother to generate the test vectors may also have a large effect on the efficacy of the preconditioner. If not enough iterations of the smoother are used, some high frequency components of the error will remain. Conversely, too many iterations will begin to reduce the low frequency components of the error, resulting in a coarse grid matrix that does not capture those parts of the error. It is thus beneficial to examine the performance of the preconditioner while varying the number of setup iterations while generating the test vectors. 
\begin{figure}[!h]
	\begin{center}
        \includegraphics[scale=0.525]{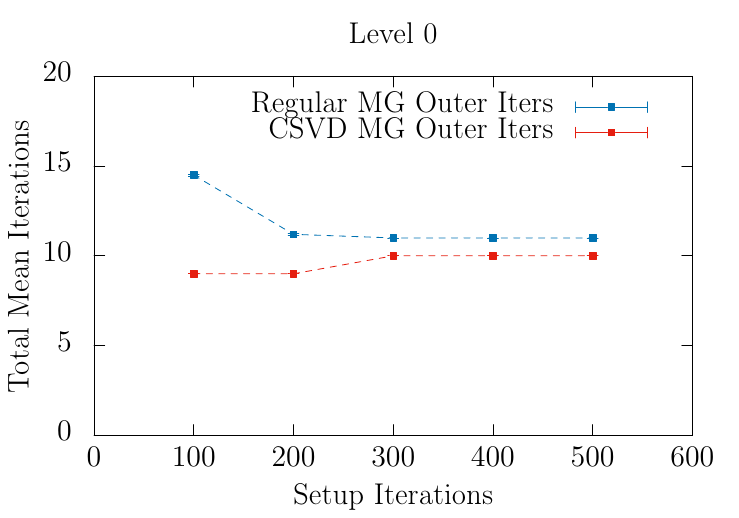}
        \includegraphics[scale=0.525]{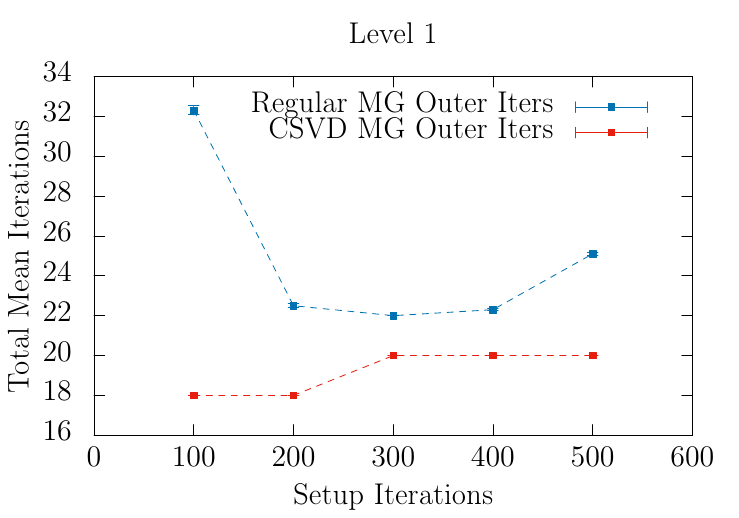}
        \includegraphics[scale=0.525]{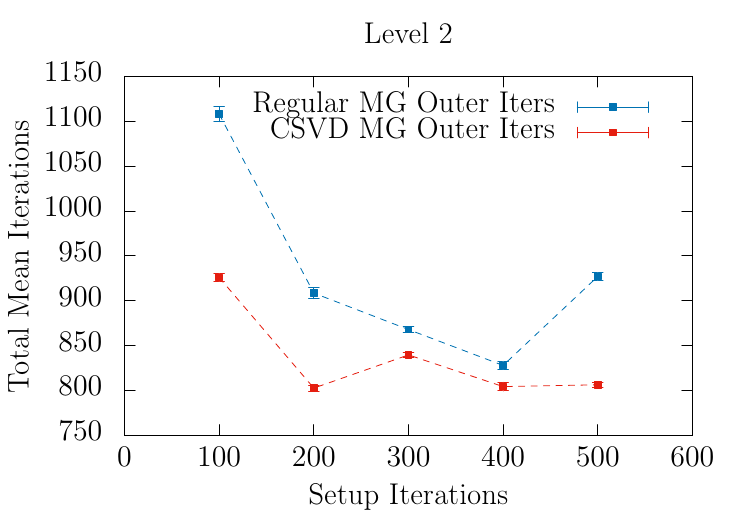}
        \includegraphics[scale=0.525]{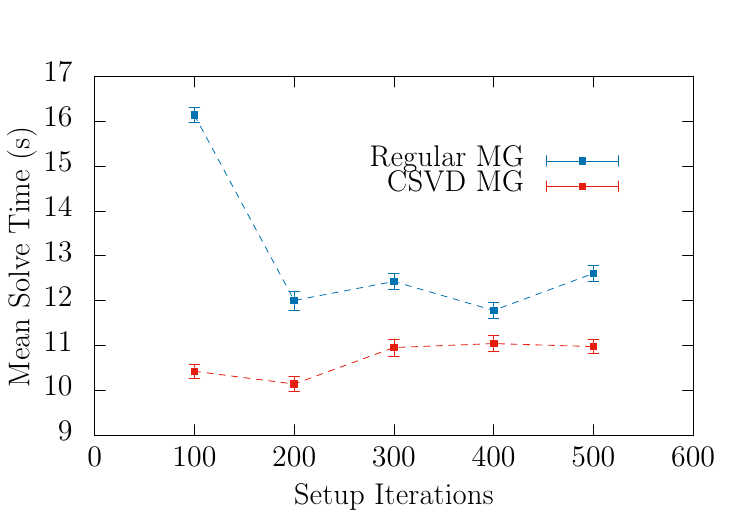}
	\caption{\label{fig::iters_aniso} The total number of iterations on $\ell = 0$ (upper left), $\ell = 1$ (upper right), $\ell = 2$ (lower left) and the mean solve time of the system of linear equations for Configuration A.}
	\end{center}
\end{figure}

\begin{figure}[!h]
	\begin{center}
        \includegraphics[scale=0.525]{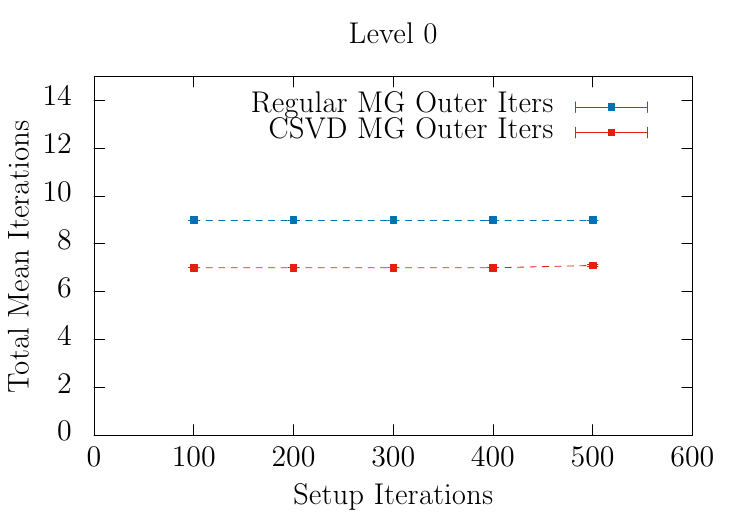}
        \includegraphics[scale=0.525]{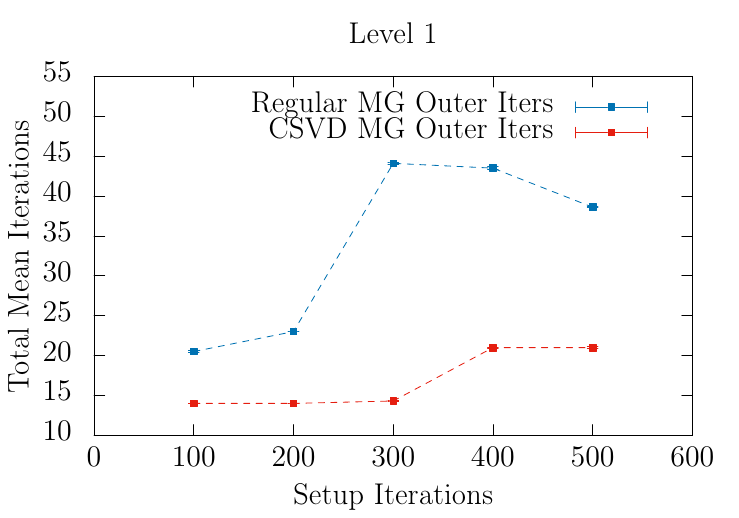}
        \includegraphics[scale=0.525]{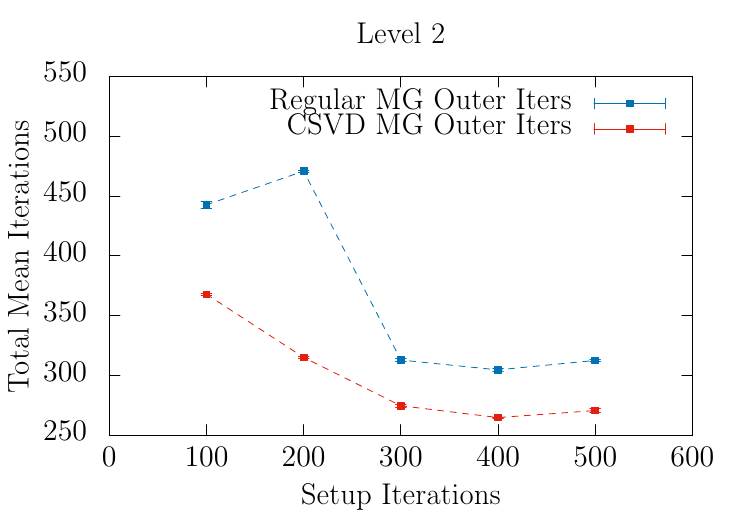}
        \includegraphics[scale=0.525]{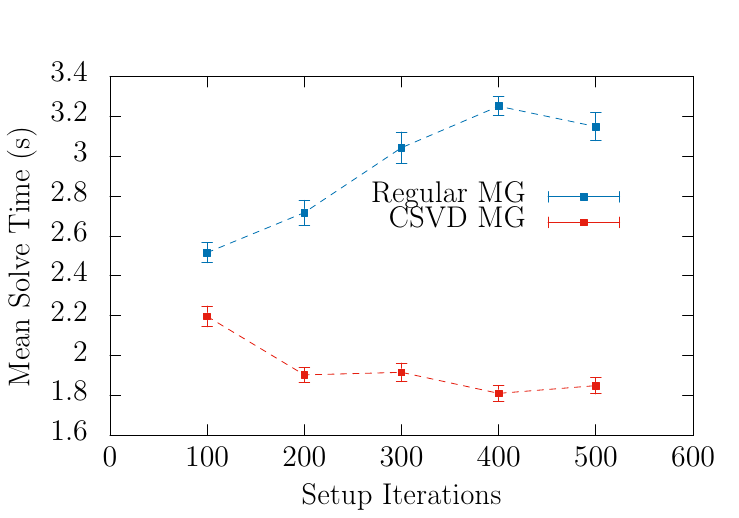}	\caption{\label{fig::iters_milc}As Figure \ref{fig::iters_aniso} for Configuration B.}
	\end{center}
\end{figure}

Figures \ref{fig::iters_aniso} and \ref{fig::iters_milc}, for configurations A and B, respectively, display the total number of iterations required at each level $\ell$ as well as the mean execution time as the number of setup iterations vary. All levels perform the same number of setup iterations and, as previously, $m = 96$ and $k = 24$ are used. For both configurations, the total cost of solving the linear system is reduced compared to conventional multigrid for all numbers of setup iterations. Additionally, the use of the CSVD results in a preconditioner that is less sensitive to the number of setup iterations. 

\begin{figure}[!h]
	\begin{center}
        \includegraphics[scale=0.525]{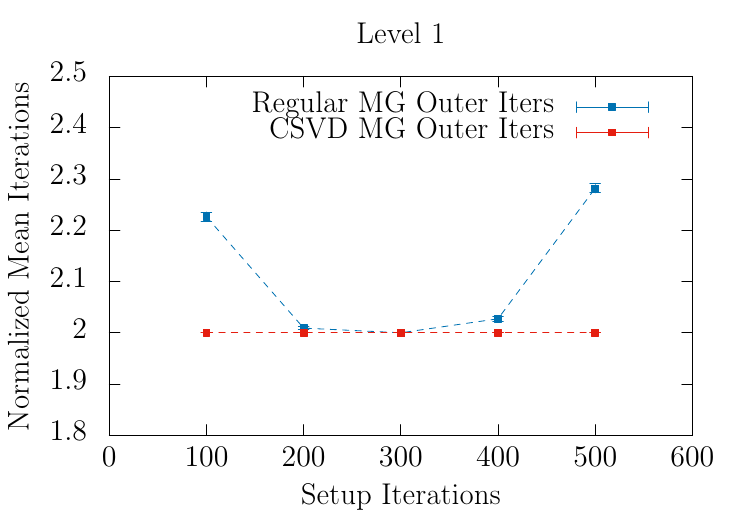}
	\includegraphics[scale=0.525]{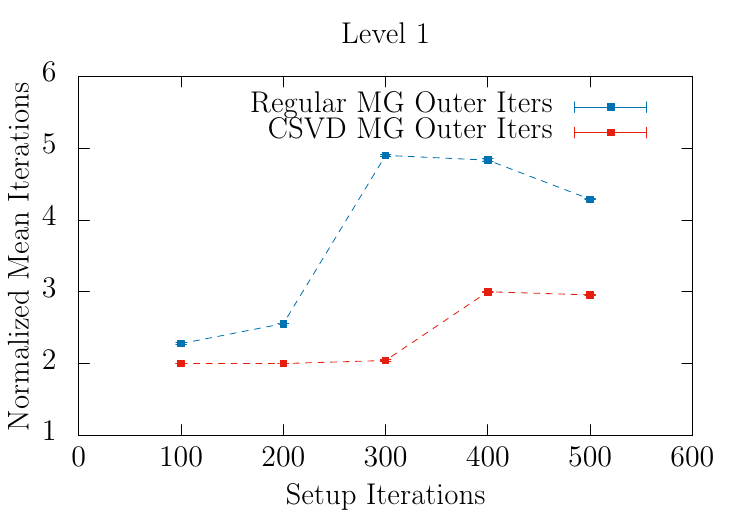}
        \includegraphics[scale=0.525]{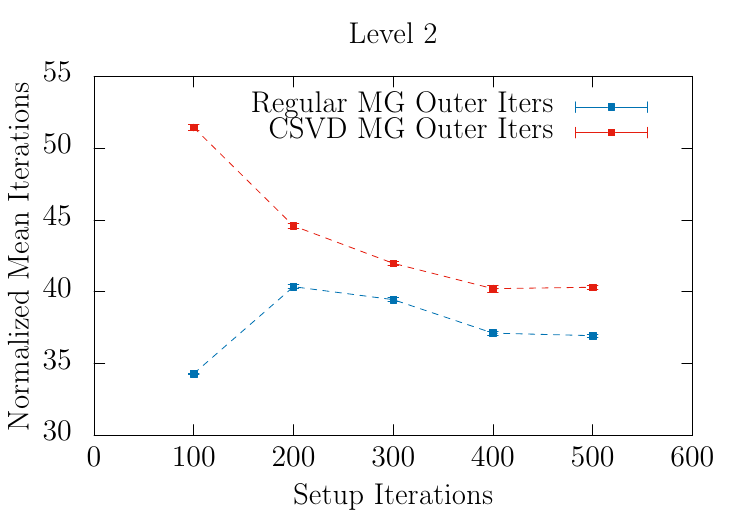}
        \includegraphics[scale=0.525]{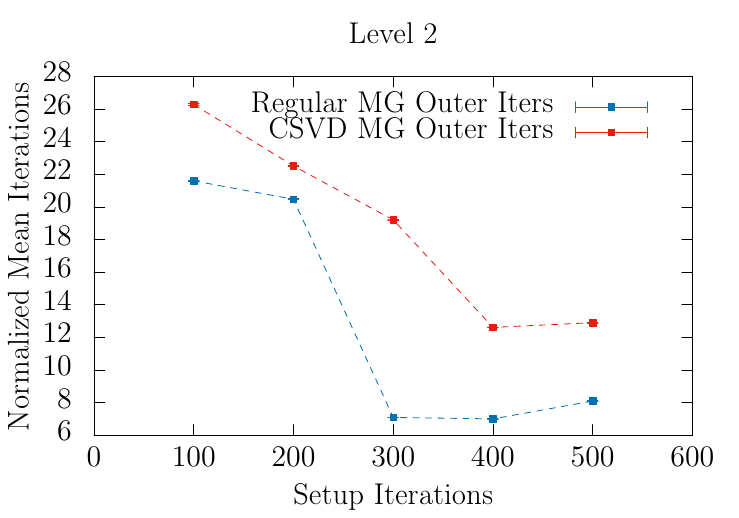}
		\caption{\label{fig::norm_iters}The normalized number of iterations on level $\ell = 1$ (top) and $\ell = 2$ (bottom) for Configuration A (left) and Configuration B (right).}
	\end{center}
\end{figure}

Besides the lower total cost of solving the linear system, it is important to quantify the performance of the preconditioner at each level per iteration due to the use of the $K$-cycle. In a $K$-cycle, multiple iterations of the solver at level $\ell+1$ occur for each iteration at level $\ell$. To examine the cost per iteration on each level, the number of iterations at level $\ell+1$ is normalized by the number of iterations on level $\ell$. Figure \ref{fig::norm_iters} displays the normalized cost for levels $\ell = 1,2$ on both configurations. It is observed for both configurations across all number of setup iterations that the cost per iteration at the coarsest level is greater for the preconditioner that utilizes the CSVD. This is consistent with expectations that more information about the low frequency components of the error is being transferred to the coarse grids where more iterations can be performed at significantly lower cost.

\subsection{Lattice Volume Scaling}
We now examine the effectiveness of the CSVD method as the volume of the hypercubic lattice increases. Systems of linear equations are solved on three lattice volumes $L_s^3\times L_t$, of spatial extent $L_s = 24,32,40$ and temporal extent $L_t = 64$, and $m_{\pi} \approx$ 220 MeV with the Clover on HISQ action\footnote{The same lattice of $L_s = 32$ is used here as in the preceding numerical results.}. The parameters of the calculation were chosen from the near-optimal setup parameters and used across all three volumes. 
\begin{figure}[!h]
	\begin{center}
	\includegraphics[scale=0.75]{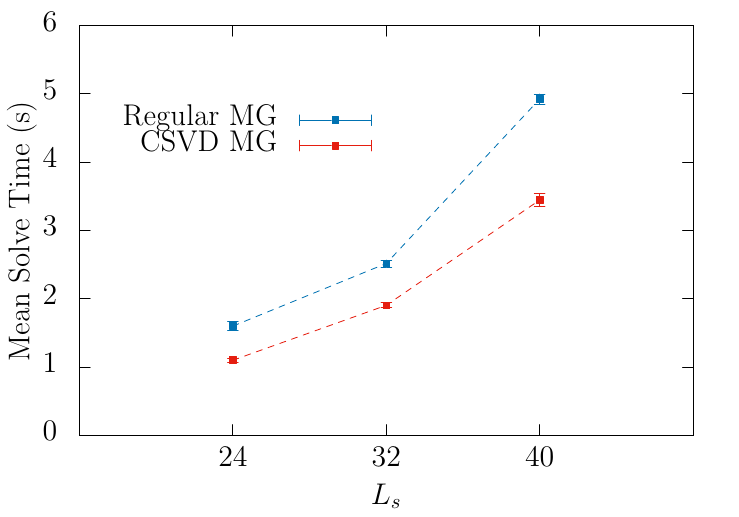}
	\label{fig::volume}
	\caption{\label{fig::volume}The performance of both MG preconditioners as the lattice volume is increased.}
	\end{center}
\end{figure}
To examine volume scaling, the mean execution time of the multigrid method is measured both with the CSVD method and without across the three volumes and reported in Figure \ref{fig::volume}. For all three volumes, the preconditioner using CSVD shows a decrease in execution time. The speedup is found to be constant across all three volumes, indicating that the CSVD method maintains its effectiveness for larger volumes.

\subsection{Streaming CSVD}
\label{subsec:stream}

We now turn to examining the effectiveness of the preconditioner generated with iCSVD method. For a direct comparison between the CSVD and iCSVD setup methods, we use a comparable number of test vectors between the two methods. The results of the previous sections showed an optimality with the use of $m = 96,128$ initial test vectors, and a truncation to the first $k=24,32$ left singular vectors for both Configuration A and B. In the streaming version of the method, this corresponds to $m = 48, 64$ with $k= 24,48$ and $r = 24,32$ for $n_s = 4$ streams. The total number of sampled test vectors is then $96$ and $128$ on levels $\ell = 0$ and $\ell = 1$, respectively, for both methods.

\begin{table}[]

\centering

\begin{tabular}{crrr|rrr}
    & \multicolumn{3}{c}{Configuration A} & \multicolumn{3}{c}{Configuration B} \\ \cline{2-4} \cline{5-7}
Level & Regular & CSVD & iCSVD & Regular & CSVD & iCSVD \\ 
    \hline
$\ell = 0$ & 11.8 & 9.0 & 9.0 & 9.0 & 7.0 & 7.0 \\
$\ell = 1$ & 23.6 & 18.0 & 18.0 & 22.7 & 14.0 & 14.0 \\
$\ell = 2$ & 973.1 & 803.4 & 801.0 & 508.1 & 331.1 & 331.0 \\

\end{tabular}
\caption{\label{tab::reg_mass}. The mean number of iterations spent on each multigrid level during a linear solve for all three setup methods with both Configuration A and B.}
\end{table}

Table \ref{tab::reg_mass} lists the mean number of iterations spent on each of the three multigrid levels during a linear solve for Configurations A and B, for the regular, CSVD and iCSVD setup methods. It is observed that there is no difference between the performance of the preconditioner created with CSVD and iCSVD methods, despite the iCSVD method producing singular vectors that are not as accurate. 

In order to asses the performance of the preconditioner using iCSVD on a more difficult problem, we now examine the performance of the preconditioner when the bare quark mass is decreased to the approximate critical mass for Configuration B. The approximate critical mass was obtained through a fit of the pion propagator on 10 configurations from the ensemble of Configuration B, and was found to be $m_{crit} \approx -0.0804$. At this value of the bare quark mass parameter, the condition number of the Wilson-Dirac operator drastically increases. Consequently, it was found that four multigrid levels are necessary as three levels did not provide the desired convergence for all methods. For the coarsest level, $m = 48$ with $k = 48$ and $r = 48$ were utilized. Adjusting the convergence criteria on level $\ell = 2$ to $||r|| = 0.06||b||$, as well as $||r||=0.6||b||$ for level $\ell = 3$ was also found to be necessary for the desired convergence. All other parameters remain the same as in previous experiments.

\begin{table}

\centering

\begin{tabular}{c@{\ \ \ \ \ }r@{\ \ \ \ \ }r@{\ \ \ \ \ }rrrr}
      &         &      & \multicolumn{4}{c}{iCSVD}\\ \cline{4-7}
Level & Regular & CSVD & $N_s = 4$ & $N_s = 8$ & $N_s = 12$ & $N_s = 16$ \\\hline

$\ell = 0$ & 12.0 & 8.0 & 8.0 & 7.5 & 7.3 & 7.0\\

$\ell = 1$ & 34.0 & 16.0 & 16.0 & 15.0 & 14.6 & 14.0\\

$\ell = 2$ & 115.2 & 58.1 & 58.0 & 53.6 & 50.8 & 48.7\\

$\ell = 3$ & 635.5 & 533.1 & 525.1 & 515.8 & 458.1& 458.0

\end{tabular}

\caption{\label{tab::mcrit}The mean number of iterations spent on each multigrid level during a linear solve for the Regular, CSVD and iCSVD setup methods for Configuration B at $m_q \approx m_{crit}$.}

\end{table}

Table \ref{tab::mcrit} displays the mean number of iterations for the regular and CSVD setup methods, as well as the iCSVD method for $n_s = 4,8,12,16$. At this lower value of the quark mass, the number of outer iterations required in the linear solve has increased by 33\% for the regular method (see Figure \ref{fig::iters_milc}). In contrast, the CSVD setup method only increases by approximately 15\%. The benefits of the iCSVD are obvious at this lower value of the bare quark mass. At the largest number of streams utilized, the number of outer iterations is constant compared to the higher value of the bare quark mass utilized in Figure \ref{fig::iters_milc}. It also provides a speedup of approximately 170\% over the regular setup method at this value of the bare quark mass.

Table \ref{tab::mcrit_norm} shows the normalized mean iterations for each multigrid level performed during the solve of the linear equations. The qualitative behavior of the methods is analogous to that of Figure \ref{fig::norm_iters}. At $\ell = 1$, the number of iterations is reduced by approximately 40\%. In contrast, $\ell = 2$ displays similar performance across all methods. The largest effect is seen on the coarsest level, where the number of normalized iterations increases drastically for the methods utilizing the CSVD. This result again reinforces that the coarsest grid is capturing more components of the low lying part of the eigenspectrum, where the error corresponding to those components can be more efficiently reduced.

\begin{table}

\centering

\begin{tabular}{c@{\ \ \ \ \ }c@{\ \ \ \ \ }c@{\ \ \ \ \ }cccc}
      &         &      & \multicolumn{4}{c}{iCSVD}\\ \cline{4-7}

Level & Regular & CSVD & $N_s = 4$ & $N_s = 8$ & $N_s = 12$ & $N_s = 16$ \\\hline

$\ell = 1$ & 2.8 & 2.0 & 2.0 & 2.0 & 2.0 & 2.0\\

$\ell = 2$ & 3.4 & 3.6 & 3.6 & 3.6 & 3.5 & 3.5\\

$\ell = 3$ & 5.5 & 9.2 & 9.0 & 9.6 & 9.0 & 9.4

\end{tabular}

\caption{\label{tab::mcrit_norm}The mean normalized number of iterations spent on each mutigrid level during a linear solve for the Regular, CSVD and iCSVD setup methods for Configuration B at $m_q \approx m_{crit}$.}

\end{table}

At the largest number of streams utilized, the setup phase of iCSVD is costly. However, as simulations reach physical quark mass and the lattice volumes become larger, the cost of such a setup phase may become necessary, as demonstrated by this artificially difficult problem.

\section{Summary and Future Work}
\label{sec:summ}
We have presented a modification to the setup algorithm for the multigrid preconditioner for systems of linear equations of Wilson fermions. The CSVD method utilizes a singular value decomposition of the chiral components of the test vectors restricted to a domain of the lattice. By calculating out a large basis of test vectors, the singular value decomposition is able to truncate the basis to the fewest vectors containing the largest contribution to the low rank approximation of the basis. In all numerical experiments, the use of the CSVD method results in a decrease in the time required to solve the system of linear equations. We also address the cost of storing the enlarged basis of vectors with iCSVD, which incrementally calculates the left singular vectors used in the prolongator. The performance of the multigrid preconditioner created using this method was shown to have the same performance as the non-streaming version. In numerical tests at critical mass, this method was shown to achieve a speedup of 170\%  in the number of fine grid iterations compared to the conventional multigrid method.

A natural extension of the approach given in this study is the unification with the least squares interpolation approach \cite{doi:10.1137/090752973}, which has been used in a bootstrap AMG algorithm for the Wilson-Dirac operator in the two dimensional lattice Schwinger model \cite{doi:10.1137/130934660}. In least squares interpolation, an optimal prolongator on the domain $\Lambda_j$ is given by
\begin{equation}
    \bm{P}_j = \bm{V}_f\bm{\Omega}\bm{V}_c^{\dag}(\bm{V}_c\bm{\Omega}\bm{V}_c^{\dag})^{-1}
\end{equation}
where $\bm{V}_f$ are the fine grid vectors restricted to $\Lambda_j$ and $\bm{V_c}$ are the coarse grid vectors. The diagonal matrix $\bm{\Omega}$ contains the weights of the minimization, which are usually set to be the energy norm of the fine grid vectors. By setting $\bm{\Omega} = I$, $\bm{V}_f = \bm{U}_k\bm{\Sigma}_k\bm{V}_k^{\dag}$ and $\bm{V}_c = \bm{\Sigma}_k\bm{V}_k^{\dag}$, it can be shown that $\bm{P}_j = \bm{U}_k$, showing that the left singular vectors are the optimal prolongator for this choice of $\bm{\Omega}$. However, this is not the only possible choice, and avenues in this direction will be explored in the future. 

\section{Acknowledgments}
\noindent TW acknowledges partial funding from the Exascale Computing Project (ECP), Project Number: 17-SC-20-SC, a collaborative effort of the U.S. Department of Energy, Office of Science and the National Nuclear Security Administration, as well as from a Royal Society Research Fellowship and the U.K. Science and Technology Facilities Council (STFC) [grant numbers ST/T000694/1, ST/X000664/1]. TW also acknowledges support from Science Foundation Ireland [grant number 21/FFP-P/10186] and Deutsche Forschungsgemeinschaft (DFG, German Research Foundation) as part of the CRC 1639 NuMeriQS – project no. 511713970. AS and ER acknowledge partial support by DOE SciDAC-5 grant (DE-FOA-0002589) and AS acknowledges partial support from National Science Foundation grant IIS-2008557. Part of this work was performed using computing resources at William \& Mary with the software codes {\tt Chroma}~\cite{chroma}, {\tt QPhiX}~\cite{qphix} and {\tt MG\_PROTO}~\cite{MGProtoDownload}. We thank the MILC Collaboration and the Hadron Spectrum Collaboration (\url{www.hadspec.org}) for making the gauge configurations available to us. The gauge configurations of the Hadron Spectrum Collaboration were generated using resources awarded from the U.S. Department of Energy INCITE program at Oak Ridge National Lab, the NSF Teragrid at the Texas Advanced Computer Center and the Pittsburgh Supercomputer Center, as well as at Jefferson Lab. We also thank Steven Gottlieb and Walter Wilcox for their help in obtaining the MILC configurations.\\

\bibliography{refs}

\end{document}